\documentclass[aps,twocolumn,showpacs,prb, superscriptaddress]{revtex4}
\usepackage{amssymb, amsbsy, amsmath, latexsym, dsfont, array, layout, graphicx, mathrsfs}
\usepackage{color}
\usepackage{ulem}

\newcommand{\one}{\mathds{1}}
\newcommand{\ket}[1]{\left|{#1}\right\rangle}
\newcommand{\bra}[1]{\left\langle{#1}\right|}

\begin{document}

\title{Nearest-neighbor coupling asymmetry in the generation of cluster states}
\author{Peng Xue}
\affiliation{Department of Physics, Southeast University, Nanjing 211189, P. R. China}
\affiliation{Institute for Quantum Information Science, University of Calgary, Alberta T2N 1N4, Canada}
\author{Barry C. Sanders}
\affiliation{Institute for Quantum Information Science, University of Calgary, Alberta T2N 1N4, Canada}
\date{\today}

\begin{abstract}
We demonstrate that charge-qubit cluster state generation by capacitive coupling is anisotropic.
Specifically, horizontal vs vertical nearest-neighbor inter-qubit coupling differs in a rectangular lattice.
We show how to ameliorate this anisotropy by applying potential biases to the array of double dots.
\end{abstract}


\maketitle

\section{Introduction}
\label{sec:introduction} One-way quantum computing is a particularly
attractive model for quantum circuits because global entanglement is
accomplished in a single step, and then all subsequent quantum
computation is effected simply by sequential feedback-controlled
single-qubit measurements~\cite{RB01}. The globally entangled state
that serves as a universal substrate for quantum computation is
known as a cluster state and was originally proposed for optical
lattices~\cite{BR01} and subsequently demonstrated with
photons~\cite{WRR+05}. Efficient quantum circuits for one-way
quantum computing was proposed for solid state devices~\cite{TLHN09}.

Solid-state charge-qubit cluster states offer the exciting prospect of one-way quantum computing with
semiconductors~\cite{TLF+06,YWTN07}.
Here we show that proposals for periodic generation of charge-qubit cluster states
involving double-dot (henceforth `ddot' as this term emphasizes the single-entity nature of
the ddot structure) charge qubits
are complicated by an overlooked
inter-qubit coupling asymmetry in two dimensions.
We remedy this complication by showing that the original proposals~\cite{TLF+06,YWTN07} can be
recovered simply by applying a potential field bias.

We proceed first by establishing the second-quantized Hamiltonian
description for the array of quantum dots and then showing how the
Hamiltonian can be simplified to a first-quantized Hamiltonian over
ddot charge qubits. In the slow tunneling-rate regime, we show that
the first-quantized Hamiltonian is well approximated by the
ubiquitous Ising-like Hamiltonian. This Hamiltonian considerably
simplifies the dynamical description and shows that the necessary
bias of the ddot charge qubits to generate cluster states is
determined by the number of ddot neighbors. Thus a global bias of a
large structure will lead to periodic evolution of excellent
approximations to cluster states. Furthermore, by applying different
biases only to the ddots on the boundary (and not to the ddots
within), the Hamiltonian induces evolution to the ideal cluster
state.

Our aim here is to remedy the deficiency of anisotropies in the
evolution of ddot charge-qubits. We show that this problem can be
nearly remedied by a global field bias and completely remedied by a
global field bias with corrective biases applied to the ddots at the
boundary. This method for correcting anisotropy is examined
numerically for the case of $40$nm GaAs ddots in a two-dimensional
lattice formation with $a$ being the distance between two sites of
the ddot, $d_x$ and $d_y$ the distance between both left and right
sites of two nearest-neighbor ddots in $x$~direction and
$y$~direction in Fig.~\ref{fig:2D}(a), respectively.

\section{Charge-qubit cluster state}
\label{sec:cqcs}
\begin{figure}
\includegraphics[width=8.5cm]{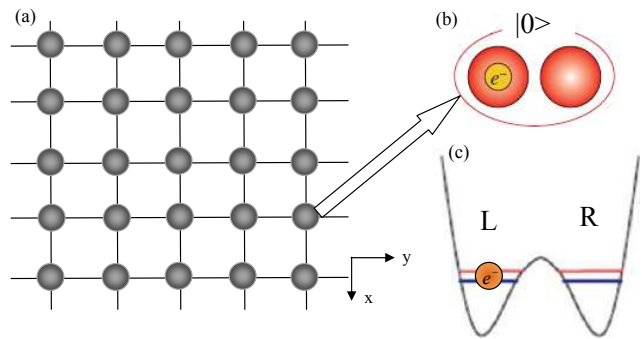}
\caption{
    (Color online.)
    (a) A two-dimensional cluster state.
    Each grey ball represents a charge qubit, and lines connect nearest neighbors in the
    horizontal and vertical directions.
    These lines correspond to coupling by controlled-Z operations.
    (b)~The qubit represented by a grey ball is expanded to a double-dot structure.
    The logical qubit state $\left|0\right\rangle$ is depicted as the double-dot structure with
    and extra charge in the left~(L) dot.
    (c)~The double-dot structure is modeled as a double-well potential.
    The blue (lower) pair of lines corresponds to the energy of the symmetric state for the sharing
    of the excess charge between the two wells,
    and the red (upper) pair of lines corresponds to the energy of the antisymmetric shared-charge state.}
   \label{fig:2D}
\end{figure}
The charge-qubit cluster state (CQCS) in two dimensions is depicted
in Fig.~\ref{fig:2D}. The standard depiction of this cluster state
is shown in Fig.~\ref{fig:2D}(a) as a periodic rectangular lattice
of qubits connected to nearest neighbors by solid lines. Each qubit
state is in $\mathcal{H}_2={\rm
span}\{\left|0\right\rangle,\left|1\right\rangle\}$, with
$\left|0\right\rangle$ the logical zero state and
$\left|1\right\rangle$ the logical one state. Cluster state
generation proceeds first by globally transforming every qubit from
the state $\left|0\right\rangle$ to the state $\left|+\right\rangle$
where
$\left|\pm\right\rangle:=\left(\left|0\right\rangle\pm\left|1\right\rangle\right)/\sqrt{2}$.
We refer to $\{\left|0\right\rangle,\left|1\right\rangle\}$ as the
`standard basis' and $\{\left|+\right\rangle,\left|-\right\rangle\}$
as the `dual basis'.

After all~$N$ qubits in the cluster state are prepared in the
$\left|+\right\rangle^{\otimes N}$ state, nearest-neighbor qubits
then interact via the two-qubit controlled-$Z$ operations,
denoted~CZ. Here~$Z$ is the Pauli `phase' operator. The other Pauli
operators are the `flip' operator~$X$, the `flip+phase' operator
$Y=XZ$, and the identity operator~$\one$. This operation is
represented in the two-qubit standard basis
$\{\left|00\right\rangle,\left|01\right\rangle,\left|10\right\rangle,\left|11\right\rangle\}$
as ${\rm CZ}={\rm diag}\left\{1,1,1,-1\right\}$. The unitary
operation~CZ is independent of whether the line in
Fig.~\ref{fig:2D}(a) is horizontal or vertical.

Fig.~\ref{fig:2D}(b) shows that the horizontal vs vertical symmetry
is in fact broken by the coupling axis of the charge qubit, which is
represented in Fig.~\ref{fig:2D}(b) as an excess charge in the left
or right quantum dot. Although the charge-qubit coupling axis could
be aligned independently of the orientation of the overall qubit
lattice, we will treat the case that the charge-qubit coupling axis
is in the $x$-direction. This is a physically reasonable case, and
extending to the case of arbitrary alignment is involved but not
difficult.

The charge qubit can be created as a semiconductor ddot structure\cite{Oo98}.
Other alternatives exist such as the superconducting charge qubit\cite{Dongen94} or a pair of dangling bonds on a surface \cite{DB}.
In any case, the logical states typically correspond either to the
left- and right-well occupancy by the excess charge or, alternatively,
to the cases of symmetric or antisymmetric charge states between the two dots of a ddot charge qubit.

For coherently evolving charge qubits, Schr\"{o}dinger's equation
can be used to describe the dynamics, and the potential in
Schr\"{o}dinger's equation is depicted in Fig.~\ref{fig:2D}(c). Here
we treat the standard basis as corresponding to left- and
right-occupancy; the dual basis then corresponds to the symmetric
and antisymmetric charge-occupancy states.

The quantum dots are engineered so that each potential well has only
one bound energy state for the excess electron. Due to the Pauli
exclusion principle, the number of electrons in each well is either
zero or one or else two electrons with opposite spins. The case of
two excess electrons in one double-dot structure should be
energetically forbidden by Coulomb repulsion between the two
electrons to preserve the integrity of the charge qubit.

\section{Modeling the dynamics}
\label{sec:modeling} The goal is to have one excess electron per
closely-spaced quantum dot pair, but the general picture is that
each quantum dot can have one or two electrons. The restriction of
one excess electron must emerge as an energetically favorable
configuration rather than be imposed by fiat. The full
second-quantized description of the electrons in the array of
quantum dots is given by the extended Hubbard model (EHM).

\subsection{The extended Hubbard model}
\label{subsec:using} The EHM applies to an array of quantum dots
whose locations in a two-dimensional array are denoted by lattice
coordinates. For $\hat{c}_{ij}$ the annihilation operator at dot
site~$(i,j)$, $\hat{c}^\dagger_{ij}$ the conjugate creation
operator, and $\hat{n}_{ij}=\hat{c}^\dagger_{ij}\hat{c}_{ij}$ the
number operator, the dynamics of the charge-qubit cluster state can
conveniently be described by the extended Hubbard
Hamiltonian~\cite{hub} (As spin is conserved, we can, without loss
of generality, assume fixed spin and ignore this degree of freedom)
\begin{align}
\label{eq:generalHam}
    \hat{H}
    =&\sum_{i,j} E\hat{n}_{ij}+\hat{V}\\
        &+\sum_{i,j,i',j'}W_{ij,i'j'}\hat{n}_{ij}\hat{n}_{i'j'}
        -T_{ij,i'j'}\left(\hat{c}^\dagger_{ij}\hat{c}_{i'j'}+\text{h.c.}\right)\nonumber.
\end{align}
Here~$E$ is the effective on-site energy for each site~$(i,j)$,
which can vary due to local field effects. $T_{ij,i'j'}$ is the
coherent tunneling rate between sites~$(i,j)$ and~$(i',j')$.
$W_{ij,i'j'}$ is the Coulomb repulsion energy between sites~$(i,j)$
and~$(i',j')$. Finally, for $\widehat{\Delta
n}_{ij,i'j'}:=\hat{n}_{ij}-\hat{n}_{i'j'}$ the number-difference
operator between sites~$(i,j)$ and~$(i',j')$, the potential bias
operator is
\begin{equation}
\label{eq:V}
    \hat{V}=\frac{1}{2}\sum_{i,j,i',j'}V_{ij,i'j'}\widehat{\Delta n}_{ij,i'j'}
\end{equation}
with~$V_{ij,i'j'}$ the inter-site $(i,j)\leftrightarrow(i',j')$
potential difference.

In fact Eq.~(\ref{eq:generalHam}) describes not just
nearest-neighbor interactions but interactions between all dots with
all other dots, where the inter-dot couplings~$T_{ij,i'j'}$
and~$W_{ij,i'j'}$ are suitably chosen. For charge qubits
corresponding to closely spaced dot pairs, $T_{ij,i'j'}$ can be
neglected for all but the ddot of a given charge qubit. Also
$W_{ij,i'j'}$ is only significant between charge qubits. The
interdot (possibly screened) Coulomb repulsion is neglected for the
ddots of a charge qubit because $W_{ij,ij}$ is sufficiently large to
prevent both dots from being simultaneously excessively charged.

\subsection{Single-qubit gates}
\label{subsec:qubits} Eq.~(\ref{eq:generalHam}) is a
second-quantized Hamiltonian. To bridge this Hamiltonian over to the
multi-qubit description, we restrict the Hilbert space, upon which
the first-quantized version of the Hamiltonian acts, to the case of
a single excess electron in each double well. Note here that a dot
has lattice coordinates expressed here as $(i_1,j_1)$ and dot~$2$ is
at $(i_2,j_2)$.

For an array with close proximity between dots of the ddot pair, the resultant charge-qubit ddot pair can be
treated as a point-like object in the quantum computing architecture.
This ddot charge qubit has a point-like coordinate designated by ~$\mathfrak{(m,n)}$
where the change of font is used to indicate point-like ddot coordinates rather than the coordinates
of a particular quantum dot.
Thus, we use $(i,j)$ to designate the location of a quantum dot in a two-dimensional array and
$\mathfrak{(m,n)}$ to denote the location of a point-like ddot charge qubit.


Assuming two dots of each ddot pair share one electron, the state od
ddot pair is a superposition of two basis states:
$\{\ket{L},\ket{R}\}$. Here $L$ ($R$) indicates that the electron is
in the left (right) dot. In this basis, $\hat{H}$ in
Eq.~(\ref{eq:generalHam}) for one ddot projected onto
$\hat{H}_\mathfrak{mn}$ is
\begin{align}
\hat{H}_{\mathfrak{mn}}
&=E(\ket{L}\bra{L}+\ket{R}\bra{R})+V(\ket{L}\bra{L}-\ket{R}\bra{R}\nonumber\\&+T(\ket{L}\bra{R}+\text{h.c.}))
=\left(
  \begin{array}{cc}
    E+V & T\\
    T & E-V \\
  \end{array}
\right)
\end{align}
for $V=V_L-V_R$ (the relative energy between the left and right dots
of a ddot pair), and~$T$ is the flip rate corresponding to the
tunneling rate between the two dots of a ddot charge qubit.

This Hamiltonian can be conveniently rewritten as a linear
combination of three types of quantum gates. These gates are the
identity~$\openone$,
$X=\left|0\right\rangle\bra{1}+\left|1\right\rangle\bra{0}$, phase
gate $Z=\left|0\right\rangle\bra{0}-\left|1\right\rangle\bra{1}$
with $\ket{0}:=\ket{L}$ and $\ket{1}:=\ket{R}$. With these
simplifications, the Hamiltonian~(\ref{eq:generalHam}) can be
projected into the qubit space. This projection becomes clear by
studying the single-qubit case comprising one ddot pair.

In the standard basis the ddot single-qubit Hamiltonian is
$\hat{H}_{\mathfrak{mn}}=E\one+TX+VZ$ at
site~$(\mathfrak{m},\mathfrak{n})$. Here $E$ is an energy term for
the qubit. The bias $V$ can be controlled by applying an electric
field potential across the ddot pair.

In order to connect our mathematical expressions to a typical
experimental setting, we consider a GaAs ddot with a single-dot
diameter of $40$nm~\cite{sci10}. A typical experimental parameter
range for tunneling is $T\approx0-10\mu$eV: here we choose
$T=0.1\mu$eV ($160$MHz). By tuning the electric field to $V=0$, the
evolution of the Hamiltonian $\hat{H}_{\mathfrak{mn}}$ effectively
implements bit flips via the~$X$ operator, and the resultant
tunneling or flipping rate is $160$MHz. By tuning the electric
potential bias to $V\gg T$, e.g.\ $V=10\mu$eV, the dynamics is
dominated by phase flipping at a rate of $16$GHz.

\subsection{Two-qubit gates}
\label{subsec:two-qubit} Now let us consider the two-qubit gate such
as the CZ gate, which can be implemented via Coulomb interaction
between two nearest-neighbor charge qubits shown below. The
Hamiltonian for two nearest-neighbor charge qubits located at
$(\mathfrak{m}, \mathfrak{n})$ and at $(\mathfrak{m'},
\mathfrak{n'})$ with (possibly screened) Coulomb interaction is
\begin{align}\label{eq:2q}
    \hat{H}_{\mathfrak{mnm'n'}}
    &=2E\one+TX_\mathfrak{mn}
        + V_\mathfrak{mn} Z_\mathfrak{mn}+TX_\mathfrak{m'n'}\\
        &+ V_\mathfrak{m'n'} Z_\mathfrak{m'n'}
        +\sum_{l,k=0}^1 \varsigma_{lk}\ket{lk}_{\mathfrak{mnm'n'}}\bra{lk}.\nonumber
\end{align} The last term describes the Coulomb interaction energy in the qubit
basis between the two nearest-neighbor charge qubits located at
$(\mathfrak{m}, \mathfrak{n})$ and at $(\mathfrak{m'},
\mathfrak{n'})$ to with sums over~$l$ and $k$ representing the two
charge qubits are in the state $\ket{l}$ and $\ket{k}$.

The coefficients $\varsigma_{lk}$ are the inter-site Coulomb
interaction strengths between the same or opposite sites of the two
charge qubits $(\mathfrak{m}, \mathfrak{n})$ and $(\mathfrak{m'},
\mathfrak{n'})$. For the case $l=k$, i.e.\ the two charge qubits are
in the same states and both electrons in the left or right dots, we
have $\varsigma_{00}=W^{\mathfrak{mnm'n'}}_{LL}$ and
$\varsigma_{11}=W^{\mathfrak{mnm'n'}}_{RR}$. For the other case
$l\neq k$, i.e.\ the two charge qubits are in different states and
the two electrons in different dots, we have
$\varsigma_{01}=W^{\mathfrak{mnm'n'}}_{LR}$ and
$\varsigma_{10}=W^{\mathfrak{mnm'n'}}_{RL}$.

In the 2D lattice shown in Fig.~\ref{fig:2D}(a), for two
nearest-neighbor charge qubits there are two cases: both qubits
located in the same column (along the $x$ direction) and in the same
row (along the $y$ direction). We use the superscripts $x$ and $y$
to distinguish these two cases. For the two nearest-neighbor charge
qubits in the same column in which the structure is symmetric,
$\varsigma_{00}^x=\varsigma_{11}^x=V_\text{Q}/d_x$, and
    $\varsigma_{01}^x=\varsigma_{10}^x=V_\text{Q}/\sqrt{d_x^2+a^2}$
for $V_\text{Q}=e^2/(4\pi\epsilon)$ with $\epsilon$ the applicable
dielectric constant. In contrast, for the two nearest-neighbor
charge qubits in the same row in which the structure is asymmetric,
$\varsigma_{00}^y=\varsigma_{11}^y=V_\text{Q}/d_y$,
    $\varsigma_{01}^y=V_\text{Q}/(d_y+a)$, and
    $\varsigma_{10}^y=V_\text{Q}/(d_y-a)$.

For the GaAs ddot system considered in the previous subsection, we have
$V_\text{Q}=1.75\times10^{-29}$N$\cdot$m$^2$. With the experimental
parameters $a=400$nm, $d_x=5.5\mu$m and $d_y=5.85\mu$m~\cite{sci10},
we numerically estimate the coefficients of the interaction terms to obtain
$\varsigma_{00}^x=\varsigma_{11}^x=20.0\mu$eV,
$\varsigma_{01}^x=\varsigma_{10}^x=19.8\mu$eV,
$\varsigma_{00}^y=\varsigma_{11}^y=18.7\mu$eV,
$\varsigma_{01}^y=17.5\mu$eV and $\varsigma_{10}^y=20.1\mu$eV.

\section{Generating cluster states}
\label{sec: generating}

In the one-way quantum computing model, the two-dimensional cluster
state is a highly entangled multi-qubit state and processed by
performing sequences of adaptive single-qubit measurements, thereby
realizing arbitrary quantum computations. The two-dimensional
cluster state serves as a universal resource for one-way quantum
computing, in the sense that any multi-qubit state can be prepared
by performing sequences of local operations on a sufficiently large
two-dimensional cluster state~\cite{NMDB06}.

In previous proposals~\cite{TLF+06,YWTN07}, charge qubits are
treated as being symmetrically coupled, which is an appropriate
strategy for the one-dimensional case but not at all for the
two-dimensional case. Here we show that, by applying local external
electric fields, we can generate two-dimensional cluster states
without the requirement of symmetry of charge qubits.

In the two-dimensional case shown in Fig.~\ref{fig:2D}(a), we obtain
a more general Hamiltonian case with external electric fields
$V_{\mathfrak{mn}}$ applied on each qubit $\mathfrak{(m,n)}$ located
in the $\mathfrak{m}^{\rm th}$ row and the $\mathfrak{n}^{\rm th}$
column of the ddot charge qubit array:
\begin{align}
    \hat{H}_\text{2D}=&N^2E
        +\sum_{\mathfrak{m,n}=1}^N \Big[TX_{\mathfrak{mn}}+V_{\mathfrak{mn}} Z_{\mathfrak{mn}} \nonumber \\
        &+\frac{1}{2}\varsigma_+^x\one_{\mathfrak{mn}}
            \otimes\one_{\mathfrak{m}+1,\mathfrak{n}}+\frac{1}{2}\varsigma_-^x
            Z_{\mathfrak{mn}}\otimes Z_{\mathfrak{m}+1,\mathfrak{n}} \nonumber\\
    &+\frac{1}{2}(\Delta\varsigma_+^x+\Delta\varsigma_-^x)Z_{\mathfrak{mn}}
    +\frac{1}{2}(\Delta\varsigma_+^x-\Delta\varsigma_-^x)Z_{\mathfrak{m}+1,\mathfrak{n}},\nonumber\\
    &+\frac{1}{2}\varsigma_+^y\one_{\mathfrak{mn}}\otimes\one_{\mathfrak{m,n}+1}+\frac{1}{2}\varsigma_-^y
        Z_{\mathfrak{mn}}\otimes Z_{\mathfrak{m,n}+1}\nonumber\\
    &+\frac{1}{2}(\Delta\varsigma_+^y+\Delta\varsigma_-^y)Z_{\mathfrak{mn}}\nonumber\\
    &+\frac{1}{2}(\Delta\varsigma_+^y-\Delta\varsigma_-^y)Z_{\mathfrak{m,n}+1}\Big],
\label{eq:2DHam}
\end{align}
where in order to rewrite the interaction terms in the Pauli operator $Z$, we introduce $2\varsigma_{\pm}^{x(y)}
        =\varsigma_{00}^{x(y)}+\varsigma_{11}^{x(y)}\pm
        \varsigma_{01}^{x(y)}+\varsigma_{10}^{x(y)}$, $\Delta\varsigma_+^{x(y)}
        =\varsigma_{00}^{x(y)}-\varsigma_{11}^{x(y)}$, and $\Delta\varsigma_-^{x(y)}=\varsigma_{01}^{x(y)}-\varsigma_{10}^{x(y)}$.

For the two nearest-neighbor charge qubits in the same column in
which the structure is symmetric, we have
$\varsigma_+^x=V_\text{Q}/d_x+V_\text{Q}/\sqrt{d_x^2+a^2}$,
$\varsigma_-^x=V_\text{Q}/d_x$ and $\Delta\varsigma_\pm^x=0$.
In contrast, for the two nearest-neighbor charge qubits in the same
row in which the structure is asymmetric, we have
\begin{equation}
    \varsigma_\pm^y=\frac{V_\text{Q}}{d_y}\pm \frac{V_\text{Q}}{2(d_y+a)}+\frac{V_\text{Q}}{2(d_y-a)},
\end{equation}
$\Delta\varsigma_+^y=0$, and
$\Delta\varsigma_-^y\equiv\Delta\varsigma=
-2V_\text{Q}a/(d^2_y-a^2)$. By choosing the proper distances $d_x$
and $d_y$ between the nearest-neighbor charge qubits, we have
$\varsigma_-^x/2=\varsigma_-^y/2\equiv\varsigma$.

For the GaAs ddot system considered in the previous section,
the energy offsets are
$\varsigma_+^x=39.7\mu$eV, $\varsigma_+^y=37.5\mu$eV, $\Delta
\varsigma=-2.6\mu$eV, and $\varsigma=10.0\mu$eV.
Note that $\Delta\varsigma$ is comparable to the bias $V_{\mathfrak{mn}}$ hence
cannot be neglected.

The Hamiltonian for the two-dimensional array of charge qubits can
be simplified as (neglecting identical terms such as $N^2E$ and
other such terms)
\begin{align}
\label{eq:hprime}
    \hat{H}'_\text{2D}
        =&\sum_{\mathfrak{m,n}=1}^N\big[T X_{\mathfrak{mn}}+V'_{\mathfrak{mn}} Z_{\mathfrak{mn}}\\
        &+\varsigma (Z_{\mathfrak{mn}}\otimes Z_{\mathfrak{m}+1,\mathfrak{n}}
            +Z_{\mathfrak{mn}}\otimes Z_{\mathfrak{m,n}+1})\big],\nonumber
\end{align}
where
\begin{equation}
\label{eq:Vprime}
    V'_{\mathfrak{mn}}=\left\{\begin{array}{ll}
        V_{\mathfrak{mn}},&\mathfrak{m}=1,\dots,N;\mathfrak{n}=2,\dots,N-1\\
        V_{\mathfrak{m}1}- \Delta \varsigma/2,&\mathfrak{m}=1,\dots,N;\mathfrak{n}=1\\
        V_{\mathfrak{m}N}+\Delta \varsigma/2,&\mathfrak{m}=1,\dots,N;\mathfrak{n}=N
        \end{array}\right.
\end{equation}
are the modified energy offsets for each ddot pair at site~$\mathfrak{(m,n)}$.

For the ddot pairs on the two edge columns, the energy offsets are
modified due to the asymmetry of the structure. In contrast, for
those in the middle columns, the energy offsets remain and are
caused by the applied electric fields because the extra energy
offsets are canceled out due to the structure.

\section{Approximating Ising-like dynamics}
\label{sec:Ising} We apply a canonical transformation for a global
basis change on the Hamiltonian shown in Eq.~(\ref{eq:hprime}):
\begin{widetext}
\begin{align}
\label{eq:effHam}
    \hat{H}_\text{eff}
        =\exp\left\{\text{i}\sum_{\mathfrak{m,n}=1}^N \frac{T}{2V'_\mathfrak{mn}}Y_{\mathfrak{mn}}\right\} \hat{H}'_\text{2D}
            \exp\left\{-\text{i}\sum_{\mathfrak{m,n}=1}^N \frac{T}{2V'_\mathfrak{mn}}Y_{\mathfrak{mn}}\right\}
    =\hat{H}_\text{Is}+\hat{H}_\text{und},
\end{align}
\end{widetext}
where
\begin{equation}
\hat{H}_\text{Is}=\sum_{\mathfrak{m,n}=1}^N\big[E_{\mathfrak{mn}}
Z_{\mathfrak{mn}} +\varsigma (Z_{\mathfrak{mn}}\otimes
Z_{\mathfrak{m}+1,\mathfrak{n}}+Z_{\mathfrak{mn}}\otimes
Z_{\mathfrak{m,n}+1})\big]
\label{eq:ising}
\end{equation} for $E_{\mathfrak{mn}}=
V'_{\mathfrak{mn}}+T^2/V'_{\mathfrak{mn}}$ an Ising-like Hamiltonian
(we need not only the interaction term such as
$Z_{\mathfrak{mn}}\otimes
Z_{\mathfrak{m}+1,\mathfrak{n}}+Z_{\mathfrak{mn}}\otimes
Z_{\mathfrak{m,n}+1}$ but also the term $Z_{\mathfrak{mn}}$ for the
generation of the cluster state) and
\begin{align}
\label{eq:uw}
&\hat{H}_\text{und}=\\&\varsigma\sum_{\mathfrak{mn}=1}^N[-\frac{T}{V'_\mathfrak{mn}}Z_\mathfrak{m+1,n}\otimes
X_\mathfrak{mn}-\frac{T}{V'_\mathfrak{m+1,n}}Z_\mathfrak{mn}\otimes
X_\mathfrak{m+1,n}\nonumber\\&+\frac{T^2}{V'_\mathfrak{mn}V'_\mathfrak{m+1,n}}X_\mathfrak{mn}\otimes
X_\mathfrak{m+1,n}+\frac{T}{V'_\mathfrak{mn}}Z_\mathfrak{m,n+1}\otimes
X_\mathfrak{mn}\nonumber\\&+\frac{T}{V'_\mathfrak{m,n+1}}Z_\mathfrak{mn}\otimes
X_\mathfrak{m,n+1}-\frac{T^2}{V'_\mathfrak{mn}V'_\mathfrak{m,n+1}}X_\mathfrak{mn}\otimes
X_\mathfrak{m,n+1}].\nonumber
\end{align} The Hamiltonian $\hat{H}_\text{und}$ is an undesirable interaction for the generation of cluster states.
In the slow-tunneling regime
\begin{equation}
\label{eq:slowtunnel}
    T\ll |\varsigma|<|V_{\mathfrak{mn}}|
\end{equation}
combined with $T\ll |V'_{\mathfrak{mn}}|$ derived from
Eqs.~(\ref{eq:Vprime}) and (\ref{eq:slowtunnel}), we obtain the
approximate Ising-like Hamiltonian $\hat{H}_\text{Is}$. If $T\ll
|V'_\mathfrak{m,n}|$ is satisfied, the coefficients of
$\hat{H}_\text{und}$ are small enough so that the unwanted
interaction term $\hat{H}_\text{und}$ can be neglected and
$E_{\mathfrak{mn}}\approx V'_{\mathfrak{mn}}$. Thus, the control
term~$V$ is now embedded within the term~$E$, which incorporates the
modified control bias~$V'$ and the tunneling rate~$T$.

\subsection{Periodically generating a cluster state}
\label{subsec:periodically}
We can periodically generate a large
particle cluster state
\begin{equation}
\ket{\tilde{\Psi}}=\exp\left\{\text{i}\sum_{\mathfrak{m,n}=1}^N
Y_{\mathfrak{mn}}\right\}\ket{\Psi}
\end{equation}
in this tilted (i.e.~biased) frame by applying the unitary operation
$\exp\left\{-\text{i}\hat{H}_\text{Is}t\right\}$ on the initial
state $\ket{\Psi}_\text{ini}$ after a time $t$, if and only if both
$\varsigma t=\frac{\pi}{4}+2k\pi$ and
$E_{\mathfrak{mn}}t=-\frac{\pi}{4}(\nu_{\mathfrak{mn}}^x+\nu_{\mathfrak{mn}}^y)+2k'\pi$
are satisfied for $k$ and $k'$ arbitrary integers. Here
$\nu_{\mathfrak{mn}}^{x(y)}$ is the number of qubits connected to
the qubit $\mathfrak{(m,n)}$ in the $x$ ($y$) axis shown in
Fig.~\ref{fig:2D}(a).

\begin{figure}
\includegraphics[width=8.5cm]{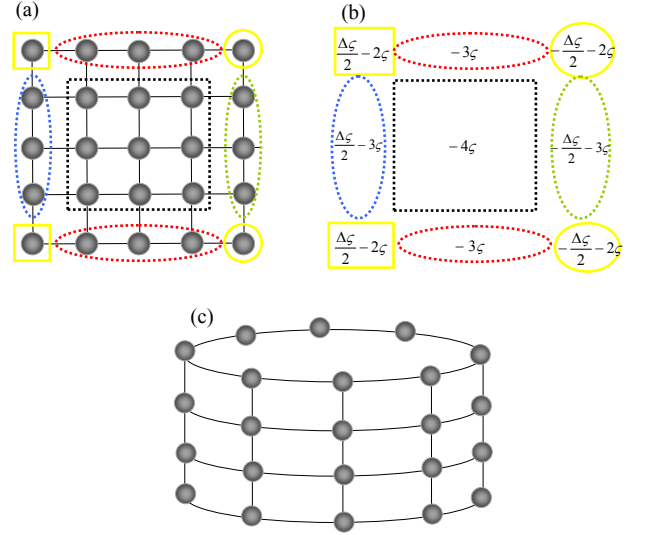}
\caption{
    (Color online.)
    (a)~The electric fields with different strengths are applied on the ddot pairs in a periodic structure isolated by the dashed and solid lines with different colors and shapes:
    $(8k'-2)\varsigma/(8k+1)+\Delta \varsigma/2$ (yellow square);
    $(8k'-2)\varsigma/(8k+1)-\Delta \varsigma/2$ (yellow circle);
    $(8k'-4)\varsigma/(8k+1)$ (black dashed box);
    $(8k'-3)\varsigma/(8k+1)$ (red dashed circle);
    $(8k'-3)\varsigma/(8k+1)+\Delta \varsigma/2$ (blue dashed circle);
    $(8k'-3)\varsigma/(8k+1)-\Delta \varsigma/2$ (green dashed circle).
    (b)~By choosing $k=k'=0$ for simplicity, we show an example of how to apply electric fields on the ddot pairs for generating a two-dimensional cluster state.
    (c)~The periodic structure of a two-dimensional array of ddot pairs for an alternative method to generate a two-dimensional cluster state by
    simply applying a global electric field $-3\varsigma$ on the first and last rows, and $-4\varsigma$ for the remaining rows, respectively.
    }
   \label{fig:electric}
\end{figure}

The two constraints lead to the relation
\begin{equation}
    E_{\mathfrak{mn}}=\varsigma\frac{-(\nu_{\mathfrak{mn}}^x+\nu_{\mathfrak{mn}}^y)+8k'}{1+8k}.
\end{equation}
Consider a two-dimensional structure shown in Fig.~\ref{fig:electric}(a): we have
\begin{equation}
\label{eq:effenergyoffsets}
    E_{11}=E_{1N}=E_{N1}=E_{NN}=(8k'-2)\varsigma/(8k+1)
\end{equation}
and
\begin{equation}
\label{eq:Emn}
    E_{\mathfrak{mn}}
        =\left\{\begin{array}{ll}
            \frac{8k'-4}{8k+1}\varsigma,&\mathfrak{m,n}=2,\dots,N-1,\\
            \frac{8k'-3}{8k+1}\varsigma, &\mathfrak{m}=1,N;\mathfrak{n}=2,\dots,N-1,\\
            \frac{8k'-3}{8k+1}\varsigma,&\mathfrak{m}=2,\dots,N-1;\mathfrak{n}=1,N.
        \end{array}\right.
\end{equation}
Equation~(\ref{eq:effenergyoffsets}) represents the effective energy
offsets of the ddot pairs at the four corners, which have two
connections along the $x$ and $y$ axes respectively; i.e.\
$\nu_x+\nu_y=2$. The similar principle applies to the first
expression of Eq.~(\ref{eq:Emn}), which shows the effective energy
offsets of the charge qubits that interact with a total of four
nearest-neighbor qubits: i.e.\ $\nu_x+\nu_y=4$. The second and third
expressions of Eq.~(\ref{eq:Emn}) show that the effective potential
differences of the ddot pairs on the boundaries with three
connections; i.e.\ $\nu_x+\nu_y=3$.

From the above reasoning we see that a cluster state can be
generated for a potential energy offset $V_{\mathfrak{mn}}$ of the
$(\mathfrak{m,n})^{\rm th}$ ddot qubit as
\begin{equation}
    V_{\mathfrak{mn}}
        \approx\left\{\begin{array}{ll}
            E_{\mathfrak{mn}},&\mathfrak{m}=1,\dots,N;\mathfrak{n}=2,\dots,N-1,\\
            E_{\mathfrak{m}1}+\Delta\varsigma/2,&\mathfrak{m}=1,\dots,N;\mathfrak{n}=1,\\
            E_{\mathfrak{m}N}-\Delta\varsigma/2,&\mathfrak{m}=1,\dots,N;\mathfrak{n}=N.
        \end{array}\right.
\end{equation}
Hence generating two-dimensional cluster states can be achieved by
applying local electric fields to set the energy offsets as above.
In our structure shown in Figs.~\ref{fig:electric}(a, b), there are
six choices of electric field biases in total with different
strengths applied to each ddot pair described in Table~I.

\begin{table}
\label{tab1}
\caption{For a ddot system in GaAs, with the choice of
$k=k'=0$ for simplicity and experimental parameters
$V_\text{Q}=1.75\times10^{-29}$N$\cdot$m$^2$, $a=400$nm,
$d_x=5.5\mu$m and $d_y=5.85\mu$m, we can estimate the potential
energy offset for each ddot as follows.}
\begin{tabular}{|c|c|c|}
  \hline
  $\mathfrak{m}$ & $\mathfrak{n}$ & $V_\mathfrak{mn}(\mu\text{eV})$ \\ \hline
  $1,N$ & $1$ & $\Delta\varsigma/2-2\varsigma=-21.3$ \\ \hline
  $1,N$ & $N$ & $-\Delta\varsigma/2-2\varsigma=-18.7$\\ \hline
  $1,N$ & $2,...,N-1$ & $-3\varsigma=-30.0$  \\ \hline
  $2,...,N-1$ & $1$ & $\Delta\varsigma/2-3\varsigma=-31.3$ \\ \hline
  $2,...,N-1$ & $1$ & $-\Delta\varsigma/2-3\varsigma=-28.7$ \\ \hline
  $2,...,N-1$ & $2,...,N-1$ & $-4\varsigma=-40.0$ \\
  \hline
\end{tabular}
\end{table}

\subsection{Cylindrical cluster state}
\label{subsec:cylindrical}
By changing the structure of the array of
charge qubits to that shown in Fig.~\ref{fig:electric}(c), we can
generate a large two-dimensional cluster state by simply applying
{\it global} electric fields instead of {\it local} electric fields,
resulting in an important simplification to the technical challenges of generating cluster states.

From our analysis above we observe that the choice of external
electric field strength depends on two factors: the number of
connections to nearest neighbors $\nu_x+\nu_y$ and the extra energy
offset due to asymmetry of the two-dimensional structure. The
periodic boundary condition $\mathfrak{n}+N=\mathfrak{n}$ dismisses
the second factor, and the extra energy offsets of the ddot pairs
located on two edge columns cancel each other.

The periodic boundary condition also partially diminishes the first
factor and makes the connections of all ddot pairs except for those
located on the first and last rows to be $\nu_x+\nu_y=4$. For those
ddots located on the first and last rows, $\nu_x+\nu_y=3$, and the
global electric fields on the two rows are set to be
$(8k'-3)\varsigma/(8k+1)$ while all the others are set to be
$(8k'-4)\varsigma/(8k+1).$ For simplicity, choosing $k=k'=0$, to
generate a two-dimensional cluster state, one only needs to apply an
external global electric field $-3\varsigma$ on the first and last
rows, and $-4\varsigma$ for the remaining rows, respectively.

\subsection{Validity of Ising-like evolution to a cluster state}
\label{subsec:fidelity} Our approach is valid if the contribution of
the term $\hat{H}_\text{und}$ in Eq.~(\ref{eq:uw}) to the evolution
is negligible. The fidelity of the pure two-dimensional cluster
state is
\begin{align}
\label{eq:fidelity}
    F   =&\left|_\text{ini}\bra{\Psi}{\rm e}^{\text{i}\hat{H}_\text{eff}t}
        {\rm e}^{-\text{i}\hat{H}_\text{Is}t}\ket{\Psi}_\text{ini}\right|^2 \nonumber \\
        \approx & 1-\left(\frac{4N\varsigma tT}{\bar{V}}\right)^2,
\end{align}
for~$N$ qubits, which gives an upper bound of the maximum number of
cluster qubits with a fixed fidelity
\begin{equation}
    N_\text{max}(N^2)=(1-F)\left(\frac{\bar{V}}{4\varsigma t T}\right)^2
\end{equation}
for~$\bar{V}$ the absolute value of the average energy offset of
ddot pairs. In other words, approximating the anisotropic evolution
by isotropic Ising-like evolution is valid provided that the total
number of qubits does not exceed $N_\text{max}$, which depends on
the acceptable less-than-unity fidelity $F$.

With the evolution time $t=\pi/4\varsigma$, the fidelity and the
maximum number of cluster qubits can be written as $F=1-\left(\pi
NT/\bar{V}\right)^2$ and $N_\text{max}(N^2)=(1-F)\left(\bar{V}/\pi
    T\right)^2$. For $T=0.1\mu$eV, $\varsigma=10\mu$eV and $\bar{V}\approx
    4\varsigma=40\mu$eV (for large $N$), our calculations show that a 2D
    $162$-qubit cluster state with a high fidelity $F=0.99$ can be produced. For $T=1\mu$eV and
    $\varsigma=10\mu$eV, $N_\text{max}=16$ with a fidelity $F=0.9$ for the 2D cluster state.

\section{Conclusions}
\label{sec:conclusions}
We have considered deterministic unitary evolution of a cluster state in a charge-qubit structure
with charge qubits made of double-dot (ddot) structures.
Although periodic evolution into charge-qubit cluster states has been considered before,
anisotropy presents a critical yet overlooked challenge.
At first anisotropy seems to destroy the opportunity to create cluster states in this way.

We have shown how to circumvent this problem by applying electric
field biases to the ddot structures. In the slow-tunneling regime,
the effective single-quantized multi-qubit Hamiltonian can be
approximated by the Ising-like Hamiltonian. In this case, electric
field biases can overcome the challenge of anisotropy. If the
electric fields had to be tailored to each ddot charge qubit, or if
the field had to be controlled over time, the strategy would be
impractical. However, we have shown that a global field over all but
the boundary qubits, and five choices of electric field biases on
boundary qubits no matter how large the system is, entirely
eliminates the problem of anisotropy.

Remarkably, by changing the structure of the array of charge qubits,
we can generate a large two-dimensional cylindrical cluster state by
simply applying an electric field on the first and last rows, and a
different one for the remaining rows, respectively. Compared to
previous schemes, no assumption of isotropy for charge qubit
couplings is made in our procedure and in fact is shown not to be
valid for two-dimensional cluster states. We augment our theoretical
analysis of anisotropy by including numerical analysis for the case
of GaAs double dots in a two-dimensional lattice. In particular we
show that the energy offsets due to anisotropy are noneligible in
this case.

For these charge qubits to be useful, noise and decoherence need to be considered.
Also the charge qubits considered here periodically evolve into cluster states and then back to their initial states
due to the periodic nature of the unitary evolution.
Timing becomes critical in such dynamics, or else the interactions that produce the cluster states must be able to
be switched off.
Measurement-based quantum computing also becomes challenging for such periodically-occurring cluster states.
These considerations are the seeds for future study.

\acknowledgments This work has been supported by National Natural
Science Foundation of China, Grant No. 10944005, the Southeast
University Start-Up Fund, Canada's Natural Science and Engineering
Research Council, the Canadian Innovation Platform ``QuantumWorks'',
and Alberta's Informatics Circle of Research Excellence.
BCS is a Fellow of the Canadian Institute for Advanced Research.

\end{document}